\documentclass[11pt,twoside]{article}
\usepackage{epsfig,amstex,amssymb,graphics}
\graphicspath{{/home/sgi/naumann/qcd}}
\newlength{\dinwidth}
\newlength{\dinmargin}
\newcommand{\ft}{$F_2~$}
\newcommand{\LO}{leading log$(Q^2)$}
\newcommand{\NLO}{next-to-leading log$(Q^2)$}
\begin{document}
\begin{titlepage}
\begin{flushleft}
{\tt DESY 95-081    \hfill    ISSN 0418-9833} \\
{\tt hep-ex/9506001} \\
{\tt May 1995}\\
\end{flushleft}
\vspace*{4.cm}
\begin{center}
\begin{Large}
\boldmath
\bf{The Gluon Density of the Proton at Low $x$\\
 from a QCD Analysis of \ft \\}
\unboldmath
\vspace*{2.cm}

H1 Collaboration \\
\end{Large}
\vspace*{4.cm}
{\bf Abstract:}
\begin{quotation}
We present a QCD analysis
of the proton structure function \ft measured by the
H1 experiment at HERA, combined with data from previous fixed
target experiments.
The gluon density is extracted from the scaling
violations of $F_2$ in the range
$2\cdot 10^{-4}<x<3\cdot 10^{-2}$
and compared with an approximate solution of the
QCD evolution equations.
The gluon density is found to rise steeply with decreasing $x$.
\end{quotation}

\vfill
\clearpage
\end{center}
\end{titlepage}
\noindent
S.~Aid$^{13}$,                   
 V.~Andreev$^{24}$,               
 B.~Andrieu$^{28}$,               
 R.-D.~Appuhn$^{11}$,             
 M.~Arpagaus$^{36}$,              
 A.~Babaev$^{26}$,                
 J.~Baehr$^{35}$,                 
 J.~B\'an$^{17}$,                 
 Y.~Ban$^{27}$,                   
 P.~Baranov$^{24}$
% ...

\noindent
% ...
$ ^{11}$ DESY, Hamburg, Germany$^a$ \\
% ...
$ ^{13}$ II. Institut f\"ur Experimentalphysik, Universit\"at Hamburg,
                                     Hamburg, Germany$^ a$  \\
% ...
$ ^{17}$ Institute of Experimental Physics, Slovak Academy of
                Sciences, Ko\v{s}ice, Slovak Republic$^ f$\\
% ...
$ ^{24}$ Lebedev Physical Institute, Moscow, Russia$^ f$ \\
% ...
$ ^{26}$ Max-Planck-Institut f\"ur Physik,
                                            M\"unchen, Germany$^ a$\\
 $ ^{27}$ LAL, Universit\'{e} de Paris-Sud, IN2P3-CNRS,
                            Orsay, France\\
 $ ^{28}$ LPNHE, Ecole Polytechnique, IN2P3-CNRS,
                             Palaiseau, France \\
% ...
$ ^{35}$ DESY, Institut f\"ur Hochenergiephysik,
                              Zeuthen, Germany$^ a$\\
 $ ^{36}$ Institut f\"ur Teilchenphysik,
          ETH, Z\"urich, Switzerland$^ i$\\
% ...
$ ^a$ Supported by the Bundesministerium f\"ur
                                  Forschung und Technologie, FRG
 under contract numbers 6AC17P, 6AC47P, 6DO57I, 6HH17P, 6HH27I, 6HD17I,
 6HD27I, 6KI17P, 6MP17I, and 6WT87P \\
% ...
$ ^f$ Supported by the Deutsche Forschungsgemeinschaft\\
% ...
$ ^i$ Supported by the Swiss National Science Foundation\\

\newpage
\section{Introduction}
The study of scaling violations of the proton structure function is a
traditional method  to obtain information on the gluon density inside the
proton~\cite{bcdmsglu,qcdnmc}.
New structure function measurements made at the
electron-proton collider HERA at
a center of mass energy of
296~GeV open  a completely new kinematic region for this study.
The accessible range in the Bjorken-scaling variable
$x$ has been extended down to about $10^{-4}$, two orders of
magnitude lower than previous
fixed target experiments.
To know the gluon density in this region is particularly
interesting since here gluons are expected to  dominate the
proton structure.
Studies of parton densities can provide a sensitive
test of perturbative QCD in the small $x$ region and can reveal the onset
of new effects, such as parton density saturation in the proton.
The gluon density or, more generally, the
parton densities inside the proton must be known in order to determine the
production rates of hadronic processes which can be described by perturbative
QCD. At future high energy hadron colliders such processes involve
parton densities at
$x$ values below $10^{-3}$.

The H1 collaboration recently published a measurement of the proton
structure function $F_2(x,Q^2)$~\cite{f2h1} derived from $ep$ scattering data
taken in 1993 corresponding to a luminosity of $271~{\rm nb}^{-1}$.
This measurement confirmed with improved
significance the observation made already in 1992 by H1~\cite{h192} and
ZEUS~\cite{zeus92} that
the structure function exhibits a strong rise towards low $x$.
This rise has caused much debate as to whether it results from conventional
DGLAP QCD evolution~\cite{dglap} of the parton
densities, or whether a new regime is entered
where the dynamics is
described by the BFKL evolution equation~\cite{bfkl}.
The latter QCD evolution equation is expected to be particularly suited for
the study of the small $x$ region since it resums all leading
$\log(1/x)$ terms in the perturbative expansion.

Using global fit methods, it  will be shown that
the measured proton structure function $F_2$  can be well described
by the DGLAP evolution equations in this new
kinematic domain. Then the DGLAP equations are  used to extract the
gluon density in the proton.
Taking into account all systematic errors of the H1 \ft measurement and their
correlations
 a full error analysis of the gluon density is performed.
A similar analysis was recently made by the
ZEUS collaboration~\cite{zeus}. A hybrid fit using
the BFKL equation
for the evolution of the gluon density at small $x$ and DGLAP
equations elsewhere is also attempted and
is found to describe the data equally well.

Besides these global fit techniques, two
approximate methods are discussed which relate the partonic densities at
a given $x$ value to the local scaling violation in that region.
Finally, following the method described in \cite{ball}
the $F_2$ data is studied in the context of
double asymptotic scaling, showing that they can be described
within this framework.

This paper concentrates on the QCD analysis of the $F_2$
measurement of H1. For a description of the analysis leading to this
measurement see to~\cite{f2h1}.
The H1 detector is described in~\cite{h1dect}.

\section{Global QCD Fits}
\subsection{Fits with the DGLAP Evolution Equations}\label{secdglap}

The DGLAP evolution equations are solved numerically
in the \NLO~ and \LO~ approximation.
Two independent programs based on the
methods of \cite{abbott} and \cite{pascaud-lal} were used
and were checked to give the same results at the percent level.
For the \NLO~ approximation
the splitting functions \cite{furman} and the strong coupling
constant $\alpha_s(Q^2)$ are defined in the $\overline{MS}$
factorization and renormalization
schemes \cite{furman2}. Starting from $Q^2_0=4~{\rm GeV}^2$,
the gluon density $g$ and the non-singlet and singlet
quark densities $q_{NS}$ and $q_{SI}$ are evolved to higher $Q^2$
values.
The singlet quark density is defined as $q_{SI}=u+\bar{u}+d+\bar{d}+s+\bar{s}$.
The non-singlet quark density is given by $q_{NS}=u+\bar{u}-q_{SI}/3$.
The following functional forms are assumed at $Q^2_0$~:
\begin{eqnarray}\label{input1}
xg(x)&=&A_gx^{B_g}(1-x)^{C_g}\nonumber \\
xq_{NS}(x)&=&A_{NS}x^{B_{NS}}(1-x)^{C_{NS}}(1+D_{NS}x)\nonumber \\
xq_{SI}(x)&=&A_{SI} x^{B_{SI} }(1-x)^{C_{SI}}(1+D_{SI}x).
\end{eqnarray}
Only proton data is used in the fit.
In order to constrain the singlet contribution without including isoscalar
target data we impose the momentum fraction carried by the gluon to be
0.44~\cite{qcdnmc} at $Q_0^2=4~{\rm GeV}^2$.
The normalization parameter $A_{SI}$ of the quark density is
then constrained by
imposing
the momentum sum rule: $\int_0^1[xg(x)+xq_{SI}(x)]dx=1$.
The shape of the gluon distribution changes only  very weakly if the momentum
 fraction carried by the gluons is varied within 6\%, which corresponds to
three times its error given in~\cite{qcdnmc}.
In the DGLAP evolution equations only three active light quark
flavours are taken into account. Heavy quark contributions are dynamically
generated
using the photon-gluon fusion prescription given in~\cite{grv1,grv2},
extended to  next-to-leading order according to \cite{riemer} and
the charm mass is assumed to be
$m_c=1.5~{\rm GeV}$. In this approach
the  contribution of beauty  quarks
remains small  and can be  neglected.
The QCD mass scale
parameter $\Lambda^{(4)}_{\overline{MS}}$ for four flavours is kept
as a free parameter in the fit.
Continuity of $\alpha_s(Q^2)$ is imposed at the charm and
beauty quark mass thresholds using  the prescription in \cite{marciano}.
The small $x$ behaviour of the gluon and the singlet quark densities are kept
independent, similar to the procedure recently advocated in~\cite{pinning}.
The structure function
$F_2(x,Q^2)$ is obtained by convoluting the evolved parton densities with the
Wilson
coefficients \cite{furman2}.

In~\cite{diffrac} it was shown
that about 10\% of the events are diffractive events, i.e.\
the electron scatters on a component of the proton which is not
colour connected with the rest of the proton.
At this stage there is no
experimental evidence that the QCD evolution of the diffractive part of
$F_2$ is significantly
different from that of the total inclusive $F_2$. Furthermore, these
contributions are most likely also present in the lower energy
data and were not subtracted there either.
Hence these events are kept in the  data
allowing a comparison
with results from other analyses.


\subsection{Fits with the Mixed DGLAP-BFKL Evolution Equations}

The BFKL equation accounts for the leading $\log(1/x)$ terms
in the perturbative expansion,
which dominate in the limit of small $x$.
Since it only describes the evolution of the
gluon density
a prescription~\cite{jan} is needed to determine $F_2$.
Here, the method detailed
in~\cite{bfkl-sok} is used. The $x-Q^2$ plane is divided into two regions~:
for $x>x_0$ the conventional \LO~DGLAP equations are used,
while
for $x<x_0$ the gluon density is extracted applying the BFKL
evolution equation. In the latter low $x$
region, the singlet quark density is computed using the \LO~DGLAP
equation, but including
the results of the evolved low $x$ gluon distribution.
The transition point $x_0$ is left as a free parameter in the fit.
Since the BFKL equation is
an evolution equation in $\log(1/x)$, an input gluon density is required
at the transition point $x_0$ for all $Q^2$ values.
This is obtained by requiring
continuity of the gluon density and of its $\log(Q^2)$ derivative at
$x_0$.
The continuity of all parton densities is preserved by taking
the same input functions (eq.~\ref{input1})
at $Q^2_0$ to solve the DGLAP equations in both regions.
For the starting gluon density at a low
$Q^2\lesssim 1~{\rm GeV}^2$
in the non-perturbative region the
prescription of~\cite{bfkl-nonpert} is used~:
$\partial xg(x,Q^2)/\partial\log(Q^2) \sim Q^2/(Q^2+k_a^2)$
where $k_a^2$ is a free parameter.
The heavy quark contribution is treated in the same way as for the full
DGLAP fit.

Based on this procedure a \LO~fit program \cite{pascaud-lal} has been written.
In order to stay in the region where $\log(1/x)$ dominates over $\log(Q^2)$
only data with $Q^2 \le 50~{\rm GeV}^2$ are used for this study.
The best fit yields a value of $x_0=10^{-2}$.

\subsection{Results of the Global Fits}

The structure function data reported by the H1 experiment cover a large
span in $x$, reaching values down to $2\cdot10^{-4}$, but do not cover the
high $x$ region.
In order to constrain the structure function at high $x$
 data from the fixed target muon-hydrogen scattering experiments
NMC~\cite{nmc} and BCDMS~\cite{bcdms} are used.
The parton densities
are fitted with the evolution equations to the \ft data, taking into account
statistical errors only to calculate the $\chi^2$.
The normalization of each experiment is left free within the limits of
their quoted errors (H1: 4.5\%, NMC(90): 1.6\%, NMC(280): 2.6\%, BCDMS: 3\%)
\cite{f2h1,nmc,bcdms}, leading to four additional fit  parameters \cite{gulio}.
To avoid regions where target corrections
and higher twist effects~\cite{qcdbcd}
could become important data from the fixed target experiments in the ranges
$Q^2<4~{\rm GeV}^2,x<0.5$ and $Q^2<15~{\rm GeV}^2,x \ge 0.5$
were not included in the fit. The $\chi^2$ was minimized using
the MINUIT program~\cite{minuit}.

\begin{figure}[htbp]
\begin{picture}(160,160)
\put(-10,-40){\epsfig{file=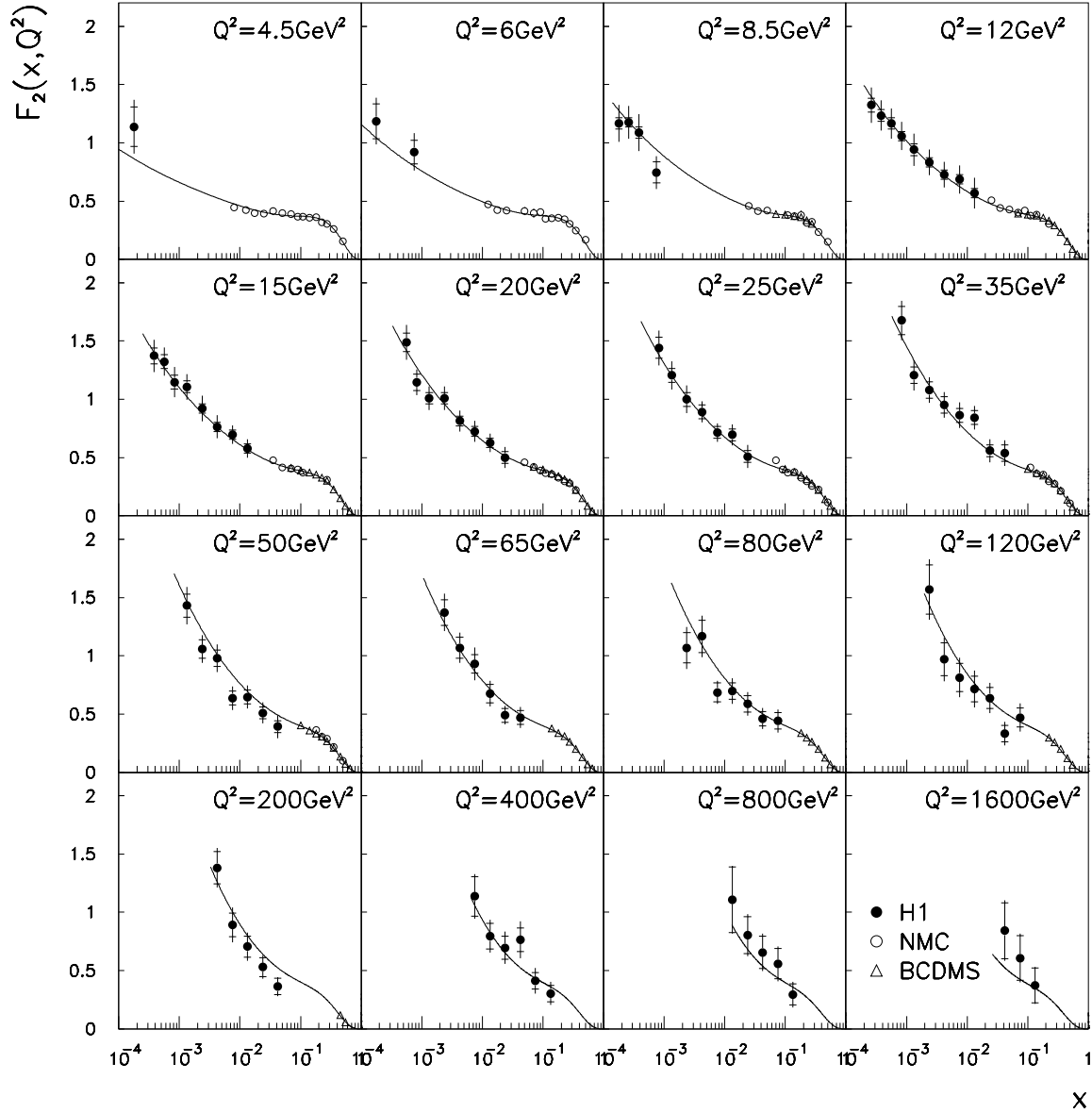,bbllx=0pt,bblly=0pt,bburx=555pt,bbury=800pt,width=16cm}}
\end{picture}
\caption{\small
$F_2(x,Q^2)$ measured by H1 together with NMC and BCDMS fixed target results.
 The inner error bar is the statistical error.
The full error represents the statistical and systematic errors added
in quadrature, not taking into account the normalization uncertainties (see
text).
The full line represents the \NLO~DGLAP fit.}
\label{f2x}
\end{figure}

\begin{figure}[htbp]
\begin{picture}(160,160)
\put(-10,-40){\epsfig{file=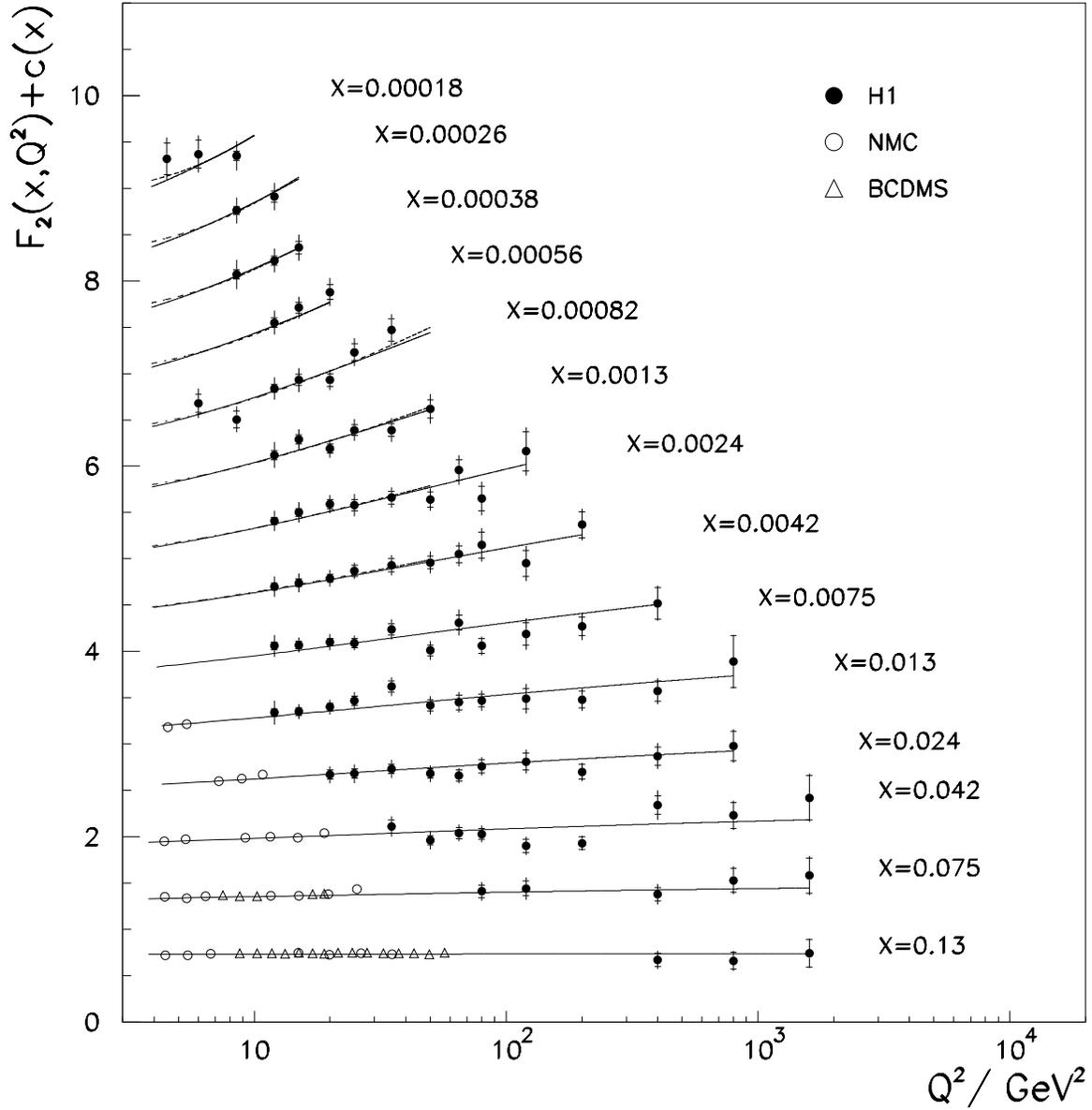,bbllx=0pt,bblly=0pt,bburx=555pt,bbury=800pt,width=16cm}}
\end{picture}
\caption[]{\small
$F_2(x,Q^2)$ measured by H1 together with NMC and BCDMS fixed target results.
The full and dashed lines represent the \LO~DGLAP and
mixed DGLAP-BFKL fits. The $F_2$ values are plotted
with a binning constant $c(x)=0.6(i-0.4)$ where $i$ is the $x$ bin number
starting
from $x=0.13$. The inner error bar is the statistical error.
The full error represents the statistical and systematic errors added
in quadrature, not taking into account the normalization uncertainties
(see text).}
\label{f2q2}
\end{figure}

The results on the fit parameters are shown in Tab.~\ref{tabpar} and
\ref{tabchi}.
The contributions to the total
$\chi^2$ from the different experiments are also given.
We obtained $\Lambda^{(4)}_{\overline{MS}}
=225~{\rm MeV}$, a value
compatible with those found by dedicated analyses \cite{booklet}.
Since the aim of this analysis is to extract the gluon density, emphasis has
 been put on the treatment of the errors on the gluon density
rather than on the fit parameters.
Fig.~\ref{f2x} shows the result of the  \NLO~QCD fit together
with $F_2$ data points as a function of $x$.
Unlike this analysis and the analysis in ~\cite{pinning}
the parameters $B_{SI}$ and $B_g$ were frequently taken to be equal
\cite{cteq,mrs}, motivated by the expectation that there
is a strong coupling between the
sea quark densities and the gluon density.
Therefore another fit was performed requiring $B_g=B_{SI}=B$, resulting in
a value of $B=-0.14$ and a $\chi^2$ increased by 3 units.
This increase stems mainly from the
lowest $x$ bin \ft measurement (Fig.~\ref{f2q2})
where the $\log(Q^2)$ slope is flatter than  expected by the DGLAP evolution.

Fig.~\ref{f2q2} presents the results from fits
using the \LO~DGLAP equations and the mixed DGLAP-BFKL equations,
together with the $F_2$ measurements.
Both the DGLAP and the mixed DGLAP-BFKL results give a good
description of the $Q^2$ evolution of the structure function.
The $\chi^2$ of the DGLAP fit is 557.
 Small differences between the two fits
 occur at the lowest $x$ bin which result
in a gain in $\chi^2$ of 7 units in favour of the mixed DGLAP-BFKL
 fit.
Thus both approaches describe the data with equal quality.
Higher precision and
measurements at lower values of $x$ and $Q^2$ expected from future HERA data
may allow to discriminate between the approaches.


\begin{table}
\begin{center}
\begin{tabular}{|c|c|c|c|c|c|c|c|c|c|c|c|}
\hline
parameter&$A_g$&$B_g$&$C_g$&$A_{SI}$&$B_{SI}$&$C_{SI}$&$D_{SI}$
&$A_{NS}$&$B_{NS}$&$C_{NS}$&$D_{NS}$\\
\hline
  &1.86&-0.22&7.12&1.15&-0.11&3.10&3.12&1.14&0.65&4.66&8.68\\
\hline
\end{tabular}
\caption{\small
Parameters of the \NLO~DGLAP fit.}
\label{tabpar}
\vspace{1cm}
\begin{tabular}{|c|c|c|c|c||c|}
\hline
Experiment&H1&NMC 90&NMC 280&BCDMS&total\\
\hline
 data points&93&34&53&174&354\\
$\chi^2$&129&55&127&192&509\\
normalization &0.93&1.00&1.01&0.97&\\
\hline
\end{tabular}
\caption{\small The number of data points, the $\chi^2$
and the normalization factors for each experiment.}
\label{tabchi}
\end{center}
\end{table}

\begin{figure}[htbp]
\begin{picture}(100,75)
\put(
0,-20){\epsfig{file=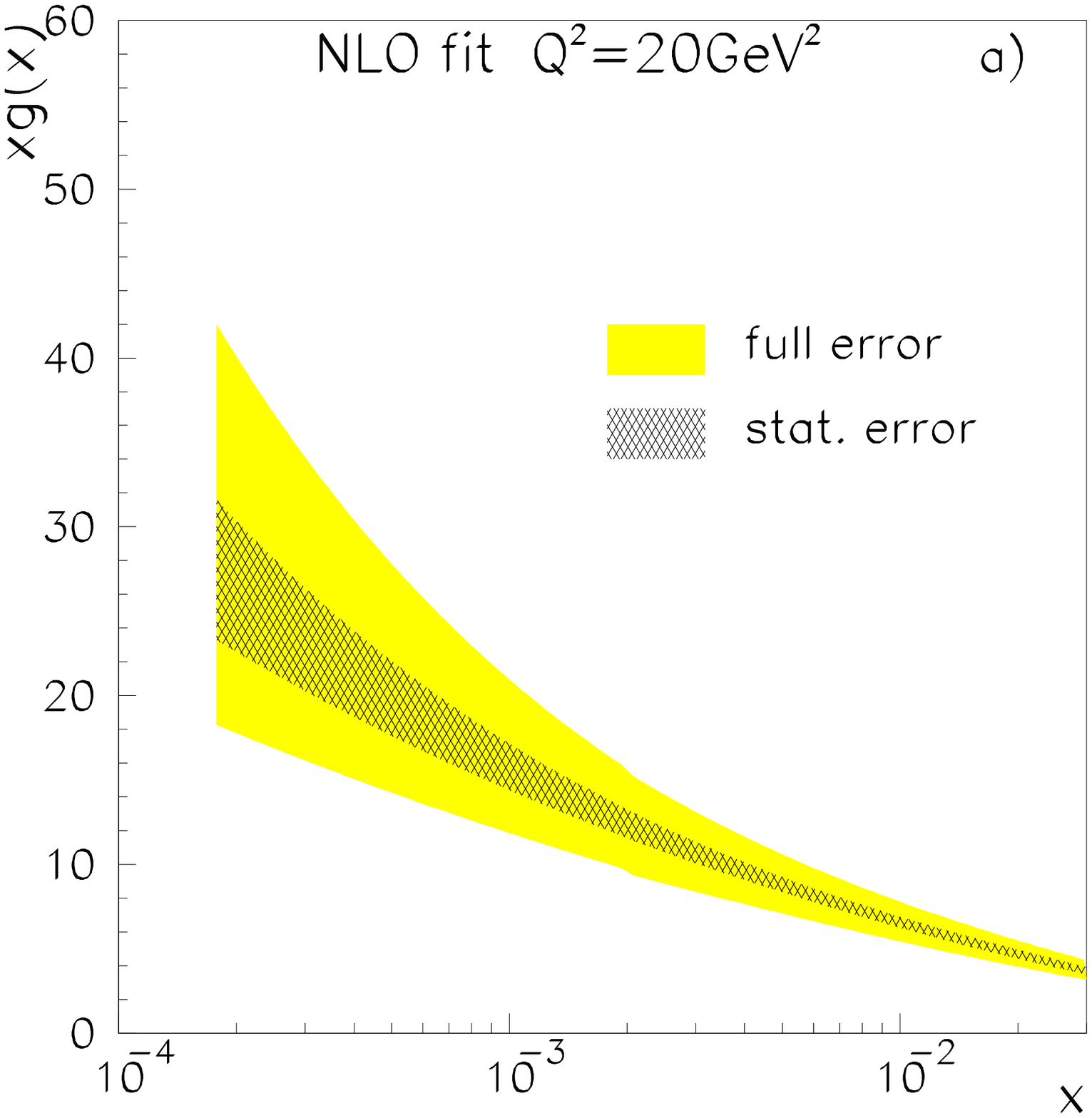,bbllx=0pt,bblly=0pt,bburx=555pt,bbury=800pt,width=7cm}}
\put(80,-20){\epsfig{file=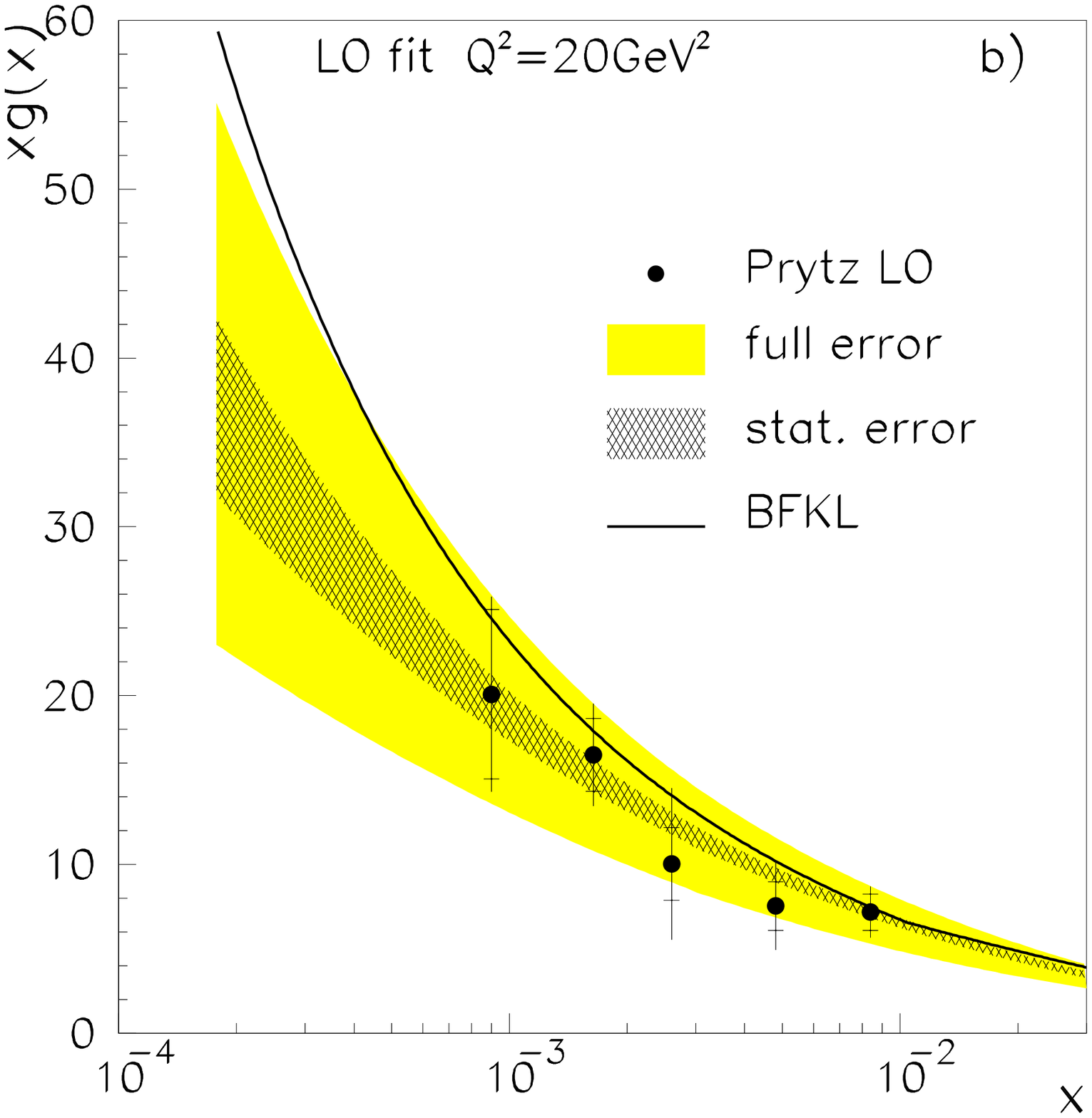,bbllx=0pt,bblly=0pt,bburx=555pt,bbury=800pt,width=7cm}}
\end{picture}
\caption[]{\small a) The gluon density $xg(x)$ at $Q^2=20~{\rm GeV}^2$
extracted from a \NLO~QCD fit.
The procedure to derive the error bands is described in the text.
b) The gluon density $xg(x)$ at $Q^2=20~{\rm GeV}^2$
from a \LO~QCD fit (shaded) and a mixed DGLAP-BFKL fit (full line).
The points are calculated using the method of Prytz.
The inner error bars represent the statistical errors.
The outer error bars are the statistical and systematic errors
added in quadrature.}\label{glu}
\end{figure}

Fig.~\ref{glu}a shows the gluon density $x g(x)$ at $Q^2=20~{\rm GeV}^2$,
extracted
with the \NLO~DGLAP evolution equations.
A strong rise of the gluon density towards low $x$ is observed.
If this rise continues with decreasing $x$, the gluons are expected to
fill up the transverse size of the proton.
A naive calculation on the limit where a uniform gluon
density fills the  proton~\cite{mueller} leads  to
$xg(x) \simeq 6 Q^2$, $Q^2$ in GeV$^2$. Hence the gluon
density determined
at $Q^2=20 ~{\rm GeV}^2$ and $x=2\cdot10^{-4}$ is still well below
 this limit.
The statistical error band of the fitted gluon results from a fit considering
the statistical errors of the data points only.

In \cite{f2h1} a full account is given of the systematic errors on
the \ft of H1.
These errors cannot be translated simply into an error on the gluon density,
since they are often strongly correlated from point to point.
A careful
study of how each individual error source affects the measured $F_2$ points
 has been made.
For example, the way in which
calibration uncertainties influence the \ft measurement depends
on the method used to reconstruct the kinematics.
{}From this study, a total of
 26 independent systematic error contributions has been established,
which cause shifts to the \ft points.
In order to properly account for the
point to point correlations a new fit parameter was introduced
for each systematic error source. The
$\chi^2$ covariance matrix and the Lagrange multiplier
method~\cite{band} were used to
calculate the error band of the gluon density shown in Fig.~\ref{glu}.

To evaluate the effect of $\Lambda$ on the gluon density
 two  fits have been performed,
fixing $\Lambda^{(4)}_{\overline{MS}}=180~{\rm MeV}$ and
$\Lambda^{(4)}_{\overline{MS}}=280~{\rm MeV}$ in order to cover the span of
values quoted in \cite{booklet}.
Further the momentum fraction carried by the gluons is varied between
 0.38 and 0.50.
The
systematic shifts between the extracted
gluon densities with different
$\Lambda^{(4)}$ and gluon momentum fraction
values are added quadratically to the systematic
uncertainty. It does not significantly affect
the total error of the gluon density in the small $x$ region.
The effect of the charm mass, which enters the
calculation via the photon-gluon fusion processes has been studied, by varying
 $m_c$
between $1~{\rm GeV}$ and $2~{\rm GeV}$. This changes the resulting gluon
density inside the statistical error band only. In the same way, the
photon-gluon fusion energy scale was changed from $2m_c$ to
$\sqrt{m_c^2+Q^2}$. The resulting
 variation of the gluon density was found to be inside
the statistical error band.
The full systematic error band of the gluon density is
shown in Fig.~\ref{glu}.

In Fig.~\ref{glu}b the gluon density resulting from
a \LO~fit of the DGLAP equations to the H1, NMC and BCDMS data is shown.
The error bands are calculated as detailed above.
The gluon density is higher compared to the \NLO~fit result.
 The value for the exponent in the gluon density
$B_g$ is $-0.32$ ($A_g=1.17$; $C_g=5.51$),
and the value for $\Lambda^{(4)}_{\rm LO}$
amounts to 185~MeV. The result of the mixed
DGLAP-BFKL fit is also shown.

\begin{figure}[htbp]
\begin{picture}(100,70)
\put(30,-20){\epsfig{file=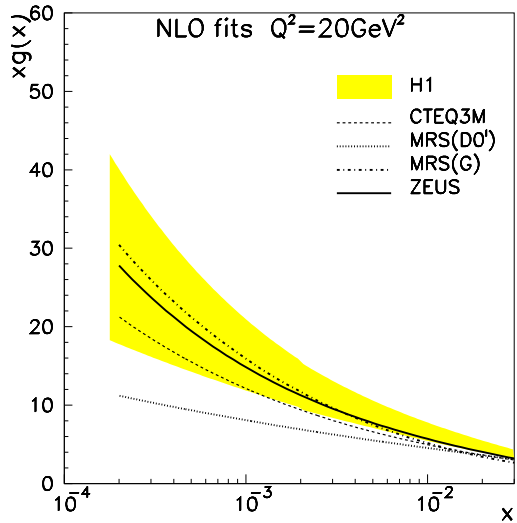,bbllx=0pt,bblly=0pt,bburx=555pt,bbury=800pt,width=7cm}}
\end{picture}
\caption[]{\small The gluon density $xg(x)$ at $Q^2=20~{\rm GeV}^2$
extracted from a \NLO~QCD fit as in Fig.~\ref{glu}a together with the ZEUS fit
{}~\cite{zeus} and the parametrizations CTEQ3M, MRSD0' and MRSG.
}\label{glupar}
\end{figure}

Fig.~\ref{glupar} shows once more the  gluon density as extracted by
H1 from the \NLO~QCD fit
together with the ZEUS fit result~\cite{zeus}
and the parametrizations of CTEQ3M~\cite{cteq}, MRSD0'~\cite{mrs}
and MRSG~\cite{pinning}.
The H1 measurement clearly disfavours the MRSD0' parametrization.
The ZEUS result and the CTEQ3M and MRSG parametrizations show a steep
rise of the gluon distribution towards small $x$ as well.
Differences of the H1 analysis with these results are due to different
data sets  used in the gluon extraction, different starting parametrizations,
a different data normalization procedure and to a smaller extent to the
different scheme used for generating heavy flavour contributions.

\boldmath
\section{Extraction of the Gluon Density from
         $\partial F_2(x,Q^2)/ \partial \log Q^2$} \label{secgluon}
\unboldmath

In this section two recently proposed approximate methods
which reduce the coupled
integro-differential DGLAP equations to simple differential equations
in $\log(Q^2)$ are discussed. The idea is to
use the logarithmic derivatives of $F_2(x,Q^2)$ locally in $x$ and $Q^2$
to approximately derive the gluon density.

In Tab.~\ref{tabdfdq}
the logarithmic derivatives
$\partial     F_2(x,Q^2)/\partial \log(Q^2)$ and
$\partial\log F_2(x,Q^2)/\partial \log(Q^2)$
 are given for different $x$ values,
determined from straight line fits to the H1 data.
These quantities are advocated to be useful for QCD studies with
approximate methods.
This usage of a straight line fit is correct to within 10\%
at $20~{\rm GeV^2}$, as derived from the phenomenological
\ft parametrization of H1 presented in \cite{f2h1}.
In order to determine the systematic errors on these values
the following procedure was followed.
All systematic errors discussed in \cite{f2h1} were used
in turn to calculate a
new set of \ft values, and new derivatives were determined by
fitting each of these new sets of shifted \ft values.
The differences of the new derivatives with the
unshifted values were added quadratically to calculate
the total systematic error.

\begin{table}
\flushleft
\begin{tabular}{|c|c|c|c|c|c|c|}
\hline
& & & & & &\\
$x$&$\frac{\partial  F_2}{\partial \log(Q^2)}$ & $\sigma_{stat}$ &
$\sigma_{syst}$
&$\frac{\partial\log(F_2)}{\partial \log (Q^2)}$ & $\sigma_{stat}$ &
$\sigma_{syst}$\\
& & & & & &\\
\hline
0.000383 & 0.51 & 0.14 & 0.09 & 0.41 & 0.11 & 0.09 \\
0.000562 & 0.65 & 0.18 & 0.10 & 0.47 & 0.13 & 0.09 \\
0.000825 & 0.46 & 0.06 & 0.06 & 0.41 & 0.05 & 0.05 \\
0.001330 & 0.28 & 0.06 & 0.11 & 0.24 & 0.05 & 0.08 \\
0.002370 & 0.21 & 0.03 & 0.06 & 0.21 & 0.03 & 0.05 \\
0.004210 & 0.20 & 0.03 & 0.03 & 0.21 & 0.03 & 0.04 \\
0.007500 & 0.08 & 0.02 & 0.03 & 0.11 & 0.03 & 0.03 \\
0.013300 & 0.06 & 0.02 & 0.02 & 0.08 & 0.03 & 0.03 \\
\hline
\end{tabular}
\vspace{-5.75cm}
\flushleft
\hspace{+9.9cm}
\begin{tabular}{|c|c|c|c|}
\hline
& & & \\
$\frac{Q^2}{{\rm GeV}^2}$ & $\frac{\partial \log(F_2)}{\partial \log(1/x)}$
&$\sigma_{stat}$ & $\sigma_{syst}$ \\
 & & & \\ \hline
8.5 & 0.19 & 0.07 & 0.06 \\
12  & 0.21 & 0.02 & 0.08 \\
15  & 0.25 & 0.02 & 0.07 \\
20  & 0.25 & 0.02 & 0.05 \\
25  & 0.28 & 0.02 & 0.07 \\
35  & 0.26 & 0.02 & 0.06 \\
50  & 0.36 & 0.03 & 0.10 \\
65  & 0.40 & 0.04 & 0.10 \\
\hline
\end{tabular}
\caption{\small The logarithmic derivatives $\partial F_2(x,Q^2)/\partial
\log (Q^2)$
and $\partial \log (F_2(x,Q^2))/\partial \log (Q^2)$ for different
values of $x$ (left), and the logarithmic derivatives $\partial \log
(F_2(x,Q^2))/\partial
log (1/x)$ for different values of $Q^2$ (right) with statistical
and systematic errors.}
\label{tabdfdq}
\end{table}

The first approximate method used was suggested by
Prytz~\cite{prytzLO}. To extract the gluon density
in \LO~he exploited the fact that
at low $x$ the scaling violations of \ft arise mainly from
pair creation of quarks from gluons.
The approximation was later
extended by the author to \NLO~\cite{prytzNLO}. Here
 only the \LO~approximation is shown since it is extremely simple.
The  assumption to neglect the quark contribution
leads to the following relation:
\begin{equation} \label{scaLO}
\frac{\partial F_2(x/2,Q^2)}{\partial\log (Q^2)}=
\frac{10}{27}\frac{\alpha_s(Q^2)}{\pi}\cdot xg(x).
\end{equation}
The result
is shown in Fig.~~\ref{glu}b together with the \LO~fit result.
The value $\Lambda=185~{\rm MeV}$ from the fit was used
as input for the Prytz approximation.
A variation of $\Lambda$ by $\pm 80$~MeV changes the result
of the approximation  by about $10\%$.
Depending on the steepness of the gluon density
for $Q^2=20~{\rm GeV}^2$ and $10^{-4} < x < 10^{-2}$
 the theoretical correction to the Prytz approximation
could rise up to --20\%, about half of which comes
from the neglect of the quark densities.
In all, the gluon density given by the \LO~fit and by the
approximate method agree within the expected precision and both show the
rise of $xg(x)$ at small $x$.

Another method was suggested by
Ellis, Kunszt and
Levin\cite{ekl} who, inspired by the BFKL equation,
assume the following shape
for the gluon and quark-singlet distributions:
$xg(x) = A_g x^{-\omega_0}$ and
$xq_{SI}(x) = A_{SI} x^{-\omega_0}$.
A prescription is given to extract
the gluon density from the
measured quantities $F_2(x,Q^2)$, $\partial F_2(x,Q^2)/\partial\log (Q^2)$
and $\omega_0=\partial\log F_2(x,Q^2)/\partial\log(1/x)$.
The slopes $\omega_0$ are determined by fitting $\log F_2$ as a function
of $\log(1/x)$. The result is shown in Tab.~\ref{tabdfdq}:
$F_2$ becomes steeper in $(1/x)$ with increasing $Q^2$.
The systematic and statistical errors on the derivatives were determined
in the same way as for the $\partial F_2(x,Q^2)/\partial\log (Q^2)$
analysis.
For $10<Q^2<50~{\rm GeV}^2$ we find $\omega_0\approx0.25$.
According to \cite{ekl},
for $\omega_0=0.25$ the approximation can only be applied
for $x<3\cdot 10^{-4}$.
Since this is essentially outside the H1 measurement region,
this method to extract the gluon density is not applied.

\section{Double Asymptotic Scaling}

Based on earlier QCD studies ~\cite{qcd}
Ball and Forte ~\cite{ball} show
that evolving a flat input distribution at $Q_0^2=1~{\rm GeV}^2$
with the DGLAP equations
 leads to a strong rise of $F_2$ at low $x$
in the region measured by HERA. An interesting feature is that if QCD
evolution is the underlying dynamics of the rise, perturbative QCD predicts
that at large $Q^2$ and small $x$
the structure function
 exhibits double scaling in the two variables:
\begin{equation}
\sigma \equiv \sqrt{\log(x_0/x)\cdot \log(t/t_0)}, \ \ \
 \rho \equiv \sqrt{\frac{\log(x_0/x)}{\log(t/t_0)}}
\end{equation}
with $t\equiv \log(Q^2/\Lambda^2)$.

This follows from a computation~\cite{qcd} of the asymptotic form of
the structure function $F_2(x,Q^2)$ at small $x$ and relies only on the
assumption that any increase in $F_2(x,Q^2)$ at small $x$ is generated
by perturbative QCD evolution.
The asymptotic behaviour of $F_2(\sigma,\rho)$ is then:
\begin{equation}
F_2(\sigma,\rho) \sim f(\frac{\gamma}{\rho})\, \frac{\gamma}{\rho}
\frac{1}{\sqrt{\gamma \rho}}\: \exp \left[ 2\gamma \sigma -
\delta \frac{\sigma}
{\rho} \right ]\; \times\; \left[ 1+{\cal O}(\frac{1}{\sigma})\right].
\label{eqbf}
\end{equation}
Here $\gamma \equiv 2\sqrt{3/b_0}$ with
$b_0$ being the leading order coefficient of the $\beta$ function
of the QCD renormalization group equation for four flavours,
$\delta = 1.36$ for four flavours and three colours. The function
$f$ depends on details of the starting distribution and tends to one
in the asymptotic limit.

In order to test this prediction
H1 data are presented in the variables $\sigma$ and $\rho$, taking
the boundary conditions to be $x_0=0.1$ and
$Q^2_0=1~{\rm GeV}^2$,  and $\Lambda^{(4)}_{\rm LO}=185~{\rm MeV}$.
\begin{figure}[htbp]
\begin{picture}(160,100)
\put(-10,-10){\epsfig{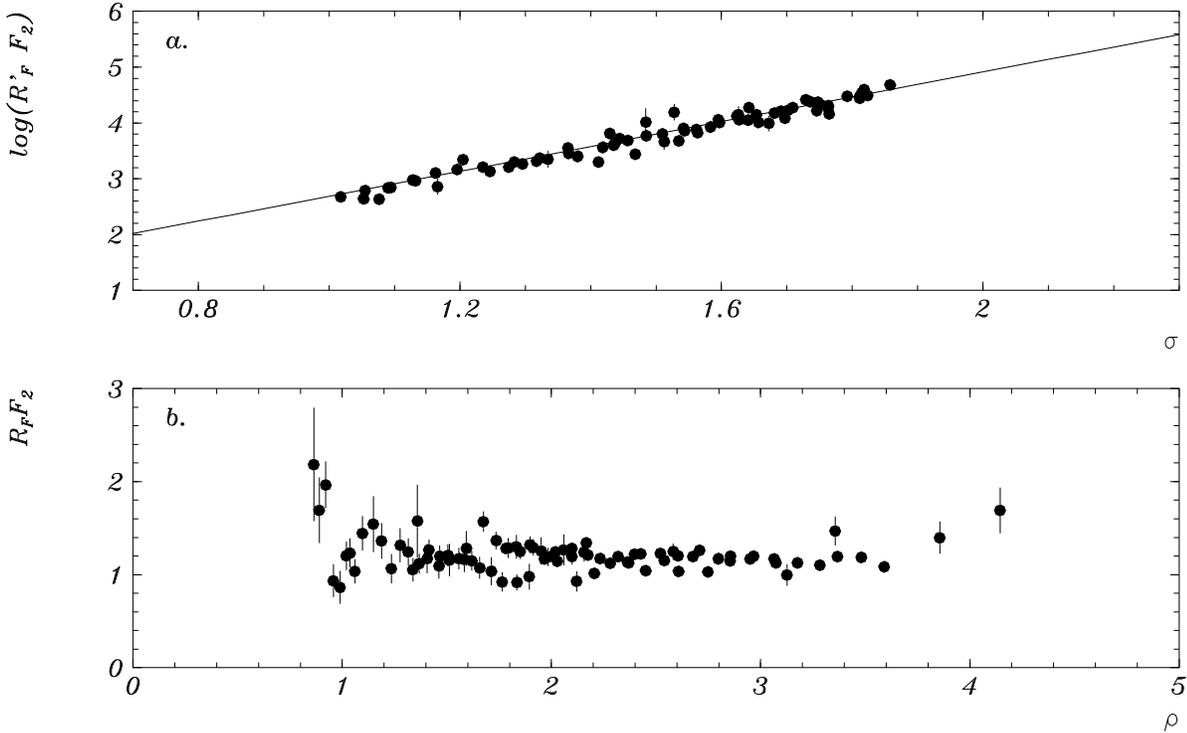}}
\end{picture}
\caption{\small  The rescaled structure functions
$\log(R'_FF_2)$ and $R_F F_2$ plotted versus the variables
$\sigma$ and  $\rho$ defined in the text. Only data with $\rho^2>1.5$ are
shown in a.}
\label{figball}
\end{figure}
The measured values of $F_2$ are rescaled by
\begin{equation}
R'_F(\sigma,\rho) = 8.1\: \exp \left(\delta \frac{\sigma}{\rho}+\frac{1}{2}
\log(\sigma) +
\log (\frac{\rho}{\gamma})\right),
\end{equation}
to remove the part of the leading subasymptotic behaviour which can be
calculated in a model independent way;
$\log(R'_F F_2)$ is then predicted to rise linearly with $\sigma$.

Fig.~\ref{figball}a shows that indeed the H1 data show a linear
rise. A fit to the data
gives a value of  $2.22\pm 0.04 \pm 0.10 $ for the slope, the first error being
statistical and  the second  systematic. The
latter was obtained in the
same way as discussed in section~\ref{secgluon}.
Varying $\Lambda$ by 80~MeV leads to an additional systematic
uncertainty of 10\%.
The result agrees well with the prediction for the slope of
$2\gamma = 2.4$, which is expected to become smaller by taking into
account higher order corrections ~\cite{ball1}.
Fig.~\ref{figball}a contains only data with $\rho^2>1.5$, for which it
is shown below that  $\rho$ is in
the asymptotic region.

Scaling in $\rho$  can  be shown by multiplying $F_2$
by the factor $R_F\equiv R'_Fe^{- 2\gamma \sigma}$ removing all the
leading behaviour in eq.~\ref{eqbf}.
This rescaled structure function should scale in both $\sigma$ and
$\rho$ when both lie in the asymptotic region:
$R_F F_2=N+{\cal O}(1/\sigma)+{\cal O}(1/\rho)$.
While the scaling in $\sigma$ can be deduced from Fig.~\ref{figball}a, we
show the
scaling  in $\rho$ in Fig.~\ref{figball}b.
Scaling sets on for
$\rho \gtrsim 1.2$ which determined the cut of $\rho^2>1.5$
for Fig.~\ref{figball}a.

The prediction for $R_F F_2$ as a function of
$\rho$
only depends on the gluon density at $Q^2_0$.
While for a soft starting gluon distribution scaling for the full asymptotic
region is predicted, a hard
gluon input would lead to scaling violations at high $\rho$~\cite{ball}.
The data shown in Fig.~\ref{figball} are well described
by the asymptotic behaviour derived from soft
boundary conditions, although
within the present precision of the data
a moderate increase at high $\rho$ is not
excluded. In addition the inclusion
of higher order corrections is expected to give a similar rise at high
$\rho$~\cite{ball1}.
\section{Summary}

QCD fits were performed  on the measured H1 proton structure function
combined with NMC and BCDMS data using both
the pure DGLAP and mixed DGLAP-BFKL evolution schemes
in the range $Q^2>4~{\rm GeV}^2$ and $2\cdot 10^{-4}<x<3\cdot 10^{-2}$.
Both prescriptions give a good description of the data.
The data do not extend to low enough $x$ or have  sufficient precision
for it to be possible to discriminate between the two approaches.
Leading $\log(Q^2)$ and \NLO~fits are
made to the data with comparable quality.

The gluon density is extracted in this region
with a full error analysis including all systematic errors.
It is found to
rise steeply with decreasing $x$.
An approximation of Prytz for the extraction of the gluon
density agrees with the QCD fit within its expected precision.
Double asymptotic scaling is observed in the region of our data.

\section*{Acknowledgements}
We are grateful to the HERA machine group whose outstanding efforts
made the structure function measurement possible. We appreciate the immense
effort of the
engineers and technicians who constructed and maintained the detector.
We thank the funding agencies for their financial support of the
experiment. We wish to thank the DESY directorate for the support
and hospitality extended to the non-DESY members of the collaboration.
We thank S.Riemersma for providing the code for the next-to-leading order
 photon-gluon
fusion model and useful discussions.
We also would like to thank J.~Kwieci\'nski, A.~Vogt, A.~Martin, R.~Roberts,
and J.~Stirling for helpful discussions.


\end{document}